\documentclass[twocolumn,english,prb,superscriptaddress]{revtex4}
\usepackage[T1]{fontenc}
\usepackage[latin9]{inputenc}
\usepackage{babel}
\usepackage{verbatim}
\usepackage{amsmath}
\usepackage{amssymb}
\usepackage{graphicx}
\usepackage{esint}
\usepackage[unicode=true,pdfusetitle,
 bookmarks=true,bookmarksnumbered=false,bookmarksopen=false,
 breaklinks=false,pdfborder={0 0 1},backref=section,colorlinks=false]
 {hyperref}

\makeatletter

\@ifundefined{textcolor}{}
{%
 \definecolor{BLACK}{gray}{0}
 \definecolor{WHITE}{gray}{1}
 \definecolor{RED}{rgb}{1,0,0}
 \definecolor{GREEN}{rgb}{0,1,0}
 \definecolor{BLUE}{rgb}{0,0,1}
 \definecolor{CYAN}{cmyk}{1,0,0,0}
 \definecolor{MAGENTA}{cmyk}{0,1,0,0}
 \definecolor{YELLOW}{cmyk}{0,0,1,0}
 }

\@ifundefined{definecolor}
 {\usepackage{color}}{}

\usepackage{times}\usepackage{mathrsfs}\usepackage{paralist}\usepackage{ifthen}\usepackage{dsfont}\@ifundefined{definecolor}{\usepackage{color}}{}

\def\be{\begin{equation}}
\def\ee{\end{equation}}
\def\bea{\begin{eqnarray}}
\def\eea{\end{eqnarray}}

\makeatother

\begin{document}

\title{Observable NMR signal from circulating current order in YBCO}
\author{Samuel Lederer and Steven A. Kivelson}

\affiliation{Department of Physics, Stanford University, Stanford, California 94305, USA}

\date{\today }
\begin{abstract}
Assuming, as suggested by recent neutron scattering experiments, that a broken symmetry state with orbital current order occurs in the pseudo-gap phase of the cuprate superconductors, we show that there must be associated equilibrium magnetic fields at various atomic sites in the unit cell, which should be detectable by NMR experiments. 
\end{abstract}
\maketitle

\section{Introduction}
	Varma\cite{Varma1997} has proposed that the pseudogap state of the high temperature superconducting cuprates is a broken symmetry phase with equilibrium circulating currents within the unit cell as shown schematically in Figure \ref{fig:cartoon}.  Indeed, recent polarized neutron scattering data\cite{Mook2008,Li2008} suggest that in at least some of the cuprates	there is a phase  transition at a pseudo-gap temperature, $T^\star>T_c$, to a state with a form of magnetic order which preserves the translational symmetries of the crystal.  (Here $T_c$ is the superconducting transition temperature and $T^\star$, determined from neutron scattering, is comparable to the previously determined pseudo-gap crossover temperature derived from transport and NMR studies.)
	
Specifically, the neutron scattering data suggest that there is some form of intra-unit-cell antiferromagnetic ordering in the pseudo-gap phase of underdoped YBCO and Hg1201 with ordered moments with magnitudes of order $0.1 \mu_B$, tilted away from the $c$ axis by a substantial angle (roughly $35^\circ-65^\circ$).   This is  broadly compatible with the Varma loop order or with an alternative model (of a sort proposed by Fauque {\it et al.}\cite{Fauque2006}
with ordered moments on O sites.  

Here, we determine the implications for Cu, Ba, and O NMR of the existence of magnetic symmetry breaking of the sort suggested by the neutron scattering experiments. Specifically, if we assume Varma loop order and take the simplest physically plausible model of the distribution of currents within the unit cell, we can analyze the neutron data to obtain quantitative estimates of the magnitudes of the putative orbital currents.  From these, we can compute the expected value of the associated magnetic fields at various sites in the unit cell. The resulting fields are large enough to be detectable, so this constitutes a falsifiable consequence of the model.

\section{Symmetry considerations}

We begin by briefly reviewing the symmetry considerations governing the existence of a static (thermodynamic) magnetic field, $\vec B(\vec r)$ at any given position in the unit cell of a crystal.  Specifically, if certain symmetries are preserved, it is possible to prove that the magnetic field vanishes.  If the field does not vanish by symmetry, the likely implication 
is that the field is non-zero. 

Let ${\cal G}_{\vec r}$ be the set of point group symmetry operations in the considered state of the system which leave the spatial point $\vec r$ invariant.  If there exists any group element $g \in {\cal G}_{\vec r}$, such that $g$ transforms $ \ B_a(\vec r) \to  -B_a(\vec r)$, it follows that $B_a(\vec r) = 0$.
 
Since $\vec B$ is odd under time-reversal ($T$), if time-reversal symmetry is unbroken, then $\vec B(\vec r)=0$ for all $\vec r$.  However, if time-reversal symmetry is broken, the field may still vanish at points of sufficiently high symmetry.  The Varma state breaks both time-reversal and rotation by $\pi$ ($R_{\pi}$) about a line parallel to the $c$ axis and passing through any atomic site, but preserves the product of these ($ T R_{\pi}$).   Since $TR_{\pi}$ transforms $B_c\to -B_c$, the out-of-plane component of the magnetic field must vanish at all atomic sites for both Hg-1201 and YBCO. 

Because $\vec B$ is a pseudo-vector, any component of $\vec B$ that lies in a mirror plane is odd under reflection through this plane.  Thus, if there is a symmetry with respect to a reflection plane passing through point $\vec r$, then all components of $\vec B(\vec r)$ within that plane must vanish.  
We will assume that the loop order leaves unbroken the reflection symmetry about the Cu-O plane in Hg-1201, and about the Y plane in YBCO, both of which we term $M$.  Along with $TR_{\pi}$ symmetry, this assures that there is no net ferromagnetism in the loop ordered state.  This reflection symmetry also implies that the in-plane components of the magnetic field vanish in the Cu-O plane of Hg-1201.

Finally, to simplify the discussion, we will assume the existence of an additional unbroken reflection symmetry $D$ about a diagonal plane containing the Cu, apical O, and Ba sites in both materials.  This is at best approximate in YBCO, which is orthorhombic in the normal state, and thus would be monoclinic in the Varma state.  However, since the YBCO lattice is tetragonal to within $2\%$, deviations from this symmetry are likely to be small.

In summary, if there exists  a Varma loop ordered state in  Hg-1201, the $M$ and $TR_{\pi}$ symmetries require $\vec B(\vec r)$ to vanish at the Cu and planar O sites. 
However, non-vanishing fields parallel to the Cu-O plane are permitted by symmetry at the apical O and Ba sites of Hg1201, and
at the Ba, Cu, apical O, and planar O sites of YBCO. 
The $D$ reflection symmetry further constrains the directions of the fields. For instance, in Hg-1201, the field at the apical O and Ba sites must point along a unit cell diagonal.
(Note that similar symmetry considerations applied to the d-density wave state\cite{Chakravarty} imply vanishing magnetic fields at the Cu and all the O sites in Hg-1201 and at the Cu, and apical O sites in YBCO, but permit an in-plane magnetic field at the in-plane O sites in YBCO, and an out-of-plane magnetic field at the Ba sites of both Hg-1201 and YBCO.)
To estimate of the order of magnitude of the expected fields, we consider the field at the center of a circular current loop of dipole moment $\mu$ and radius $r$:
\[
B(\mu,r)=\frac{2\mu}{cr^3}=325 G \left(\frac{\mu}{\mu_B}\right)\left(\frac{a}{r}\right)^3 ,
\]
where $a=3.8$\AA  is the 
Cu-Cu distance.  Taking $r=\frac{a}{2}$ and $\mu = 0.1\mu_B$ (as reported by Mook {\it et al.}\cite{Mook2008}),
 this
 gives $B \sim 250G$.
\section{Explicit Model}
For simplicity, we consider a model in which
 only the Cu-O planes are electronically active, and in which we ignore the orthorhombicity of the YBCO lattice and any small buckling of the in-plane O-Cu-O bond.  We assume the symmetries mentioned in the previous section as well as lattice translation symmetry.  

\begin{figure}
\includegraphics[clip=true,scale=0.5]{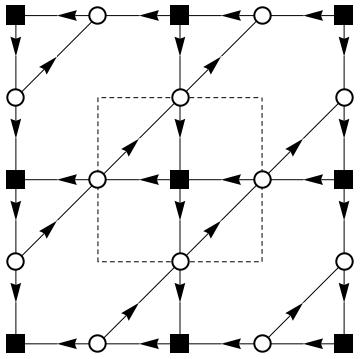}
\caption{Top view of circulating currents in the Varma loop ordered state.  Cu sites are black rectangles, O sites white circles.}
\label{fig:cartoon}
\end{figure}

\begin{figure}
\includegraphics[clip=true,trim=20 70 20 60, scale=0.7]{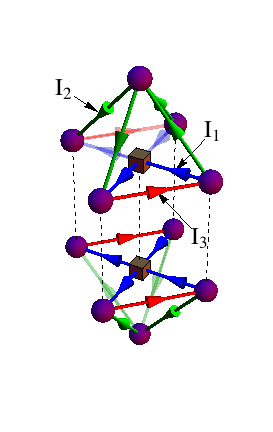}
\caption{Microscopic model for YBCO. O sites are spheres, Cu sites are cubes.}
\label{fig:YBCO}
\end{figure}

\begin{figure}
\includegraphics[clip=true,trim=20 90 20 60, scale=0.7]{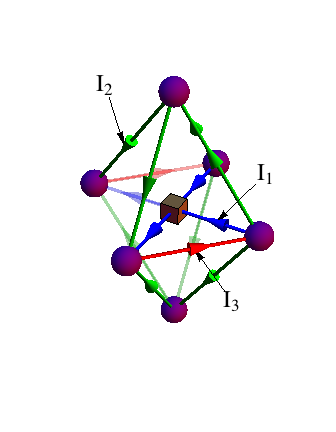}
\caption{Microscopic model for Hg1201. O sites are spheres, Cu sites are cubes.}
\label{fig:Hg1201}
\end{figure}

The ideal two dimensional cartoon of the state originally proposed by Varma is sketched in Figure \ref{fig:cartoon}.   Since in the cuprates, the distance to the apical O is comparable to the distance between planar O's, we must take into account currents involving the apical O, even if we take the simplest (shortest-range) version of this state adapted to the actual materials. (This was recognized previously in Refs. \cite{Li2008,Mook2008,Shekhter2009}, as it is necessary to account for the presence of the in-plane components of the magnetic order inferred from the neutron scattering experiments.)  The assumption of near-neighbor currents and the symmetries previously assumed lead to the pattern of currents shown in  Figures \ref{fig:YBCO} and \ref{fig:Hg1201}
 for YBCO and Hg 1201, respectively.  Note that  $TR_{\pi}$ symmetry forbids a current between the apical O and Cu sites, while the reflection symmetries $M$ and $D$ ensure that the precise pattern of broken symmetry in the proposed state is defined by the three independent currents labeled $I_j$ in the figures.

The model presented is broadly consistent with the neutron scattering data in that it would result in magnetic scattering intensity at suitable Bragg vectors with a polarization dependence reflecting comparable strengths of the in-plane and out-of-plane  magnetic fields.  
However,  there is at least one unresolved discrepancy: The in-plane component of the magnetization at $(100)$ must vanish so long as $M$ is unbroken.  Experimental data\cite{Mook2008} find a tilted moment at $(100)$, though this is only reported for one experiment in the published literature, and with a substantial margin of error.

Before making a quantitative comparison with experiment, we have imposed the additional constraint, not required by symmetry, that there be no net current flowing through the system in equilibrium.  This is equivalent to taking $I_3=I_1+I_2$ in Figures \ref{fig:YBCO} and \ref{fig:Hg1201}.  Assuming that "form factor'' effects associated with the spatial extent of these currents can be neglected, the resulting magnetic field  that would be produced by this state can be directly computed (as sketched in the Appendix) in terms of the two remaining independent parameters $I_1$ and $I_2$.  These can be determined by comparing the measured and predicted spin-flip magnetic scattering cross-sections for orthogonal neutron polarizations at a single Bragg wave-vector.  As the strongest scattering (and consequently the smallest experimental uncertainty) occurs at $(011)$ in both YBCO and Hg-1201, we determine the values of $I_j$ for each material using data from that peak only. Since the scattering cross-section is quadratic in the currents, there are two independent solutions for $I_1$ and $I_2$.Naturally, once the currents are determined, it is straightforward to compute the field at any particular spatial point.  We quote results for each of the two independent solutions for $I_1$ and $I_2$.

\section{Results and discussion}
For the case studied in Ref. \cite{Mook2008} of YB$_2$C$_3$O$_{6+\delta}$ with $\delta=0.6$ and $T_c=61K$,  
the analysis described  above leads to the prediction of fields of 
 $100 $'s of G at various sites in the unit cell, consistent with the 
dimensional estimate presented above.  In particular, we predict fields of $700 G$ or $710G$ at the copper site, $230G$ or $130G$ at the barium site, $170G$ or $280G$ at the apical oxygen sites, and $360 G$ or $350G$ at the in-plane oxygen sites.  Additional symmetries in Hg1201 lead to quite different results. 
In particular,  
as required by symmetry, the fields at both the Cu and the planar O sites vanish exactly.  However, if we apply the present analysis to the neutron data of Ref. \cite{Li2008} for underdoped Hg 1201 ($T_C=81 K$), we infer that at low temperatures there should be a field of $180G$ or $200 G$ at the apical oxygen site, and $240G$ or $170G$ at the Barium site.  Observed\cite{Takigawa} Cu NMR line-widths in YBCO are of order 100G and O linewidths substantially smaller, so these effects should be readily observable.

There are, of course, uncertainties in the quantitative estimates we have made associated with the error bars on the neutron data.  Most importantly, the assumed pattern of currents is not fully consistent with the neutron data;  as mentioned above, it follows from the assumed symmetries of the state that the in-plane moment at (100) must vanish, whereas a small, in-plane component is reported in experiment\cite{Mook2008}.  There are also theoretical uncertainties associated with the simplifying assumptions that we have made in defining the model used for explicit calculations.  Specifically, some quantitative changes could arise from form factors associated with the non-zero spatial extent of the equilibrium currents, and with currents flowing between further neighbor sites in the lattice.  We do not believe, but have not proven, that these uncertainties will not cause qualitative revisions in the quantitative estimates we have made.  We have also treated the currents classically, which is a valid procedure in typical situations involving broken symmetry states;  however, Varma\cite{He2011} has argued that there is an essentially quantum character to the loop ordered state which invalidates such an approach.  Finally, a magnetic ordering that appears static within the frequency resolution of neutron scattering may be rapidly fluctuating on NMR timescales. Though no obvious mechanism exists for an order fluctuating on intermediate timescales, it would be consistent with the neutron results and the symmetries assumed, yet yield a null result in NMR experiments.  In any case, the presence of static magnetic fields at the various atomic sites in the crystals, if seen, would constitute strong evidence of  the putative current loop order in the pseudo-gap phase of these materials.

At present, we are unable to find published Cu or O NMR studies which give clear results concerning the existence or absence of fields of the predicted magnitude in YBCO or Hg1201.  Work by Str\"{a}ssle {\it et al.}\cite{Strassle2011} sets an upper bound of less than $1 G $ for the field at the Barium site of YBCO 248.  However, since neutron scattering is not available for this material, this bound cannot be directly compared with a predicted magnitude.
We hope that the results  in this Brief Report will encourage experimental tests of local magnetic fields in the pseudogap state of the aforementioned materials.

{\bf Acknowledgements:}  We thank C. Varma and P. Bourges for helpful comments.

\section{Appendix: Sketch of Calculations}
The neutron scattering data used in these calculations are the cross sections for elastic magnetic neutron scattering in the spin-flip channel for two independent polarizations.  
The Born approximation gives the differential scattering cross section at momentum transfer $\vec K$ (a reciprocal lattice vector) as
\[
I_{\hat{ P}}(\vec K)=\left(\frac{m_N}{2\pi\hbar^2}\right)^2\left(\frac{g\mu_N}{2}\right)^2\left| \vec B_{\vec K}-\hat{ P}\left(\hat{ P}\cdot \vec B_{\vec K}\right)\right|^2
\]
where $g=- 3.826$ is the neutron g-factor, $\mu_N$ the nuclear magneton, $\hat{ P}$ the polarization direction of the incoming neutrons, and $\vec B_{\vec K}$ is the Fourier transform of the magnetic field. The momentum transfer $\vec K$ employed in Refs \cite{Mook2008,Li2008,Fauque2006} is $(0,1,1)$.  The magnetic field component at this wavevector can be expressed in terms of the currents $I_1$, $I_2$, and $I_3$ as
\[
\vec B_{\vec K} =- \frac{4 \pi i}{c}  \frac{ \vec K \times \vec J_{\vec K}}{K^2}
\]
where $\vec K$ is a reciprocal lattice vector, and the current density $\vec J_{\vec K}$ can be expressed as
\[ 
\vec J_{\vec K}=-2\sum_n I_{j_n}\hat{e}_n\frac{\sin\left(\vec K \cdot \hat{e}_n \frac{L_n}{2}\right)}{\vec K \cdot \hat{e}_n}e^{i\vec K \cdot \vec{ \cal{R}}_n}.
\]
In this expression the sum runs over the current-carrying bonds (segments) in a unit cell.  For a given segment $n$, $I_{j_n}$ is the current ($I_1$, $I_2$, or $I_3$, as defined previously), $\hat{e}_n$ is the direction of current flow, $L_n$ is the length of the bond, and $\vec{\cal R}_n$ is the position of the bond midpoint.

With these results, we can write the scattering cross section as a function of the currents, and then invert to solve for these currents in terms of the experimental data.  We then evaluate the real space magnetic field using the Biot Savart Law.  
All told, the field is
\begin{align*}
\vec B(\vec r)=\sum_{\vec R}\sum_n\frac{I_{j_n}}{c}\frac{\vec a \times \hat{e}_n}{ \left|\vec a-( \hat{e}_n\cdot \vec a)\hat{e}_n\right|^2} \\\cdot
\left(\frac{\vec a \cdot \vec{e}_n+\frac{L_n}{2}}{|\vec a+\frac{L_n}{2}\vec{e}_n|} -\frac{\vec a \cdot \vec{e}_n-\frac{L_n}{2}}{|\vec a-\frac{L_n}{2}\vec{e}_n|}\right)
\end{align*}
In this expression the first sum runs over all unit cells of the crystal (i.e. $\vec R$ runs over all Bravais lattice vectors), the second sum is over the current-carrying bonds in a unit cell as above, and the position of a given segment is $\vec a \equiv \vec{a}_{n,\vec R}=\vec R-\vec r + \vec{ \cal R}_n$.  The sum converges rapidly.

\bibstyle{plain}
\bibliography{Final_Version.bib}

\end{document}